\documentclass[amssymb,aps,prl,twocolumn,amsmath,groupedaddress]{revtex4}
\usepackage{graphicx}

\begin{document}

\title{Quantum control of optical four-wave mixing with femtosecond 
$\omega$-$3\omega$ laser pulses: coherent ac Stark nonlinear spectroscopy}
\author{Carles Serrat}
\affiliation{Departament de F\'{\i}sica i Enginyeria Nuclear, Universitat Polit\`{e}cnica de Catalunya,
Colom 1, 08222 Terrassa, Spain}

\date{\today}

\begin{abstract}
The four-wave mixing produced with two ultrashort phase-locked $\omega$-$3\omega$ laser pulses 
propagating coherently in a two-level system in the infrared spectral region is shown to depend
on the pulses relative phase. The Maxwell-Bloch equations are solved beyond the rotating-wave 
approximation to account for field frequencies which are largely detuned from the atomic resonance.
The relative phase dominating the efficiency of the coupling to the $5\omega$ anti-Stokes 
Raman component is determined by sign of the total ac Stark shift induced in the system, in such
a way that the phase influence disappears precisely where the ac Stark effect due to 
both pulses is compensated. This fundamental quantum interference effect can be the basis for
nonlinear ultrafast optical spectroscopy techniques.
\end{abstract}


\maketitle

Phenomena arising from the coherent control of nonlinear interactions are of importance 
in fields as diverse as optoelectronics and materials research, in high harmonic generation, 
in photoionization or molecular dissociation, and in biological applications such as 
spectroscopy and imaging, among others \cite{shapiro,photodiss,Watanabe,photoion,Bandrauk,Brown,Xu}.
Optical quantum coherent control is based on the fact that the phases of 
interfering transition amplitudes in light-matter interactions can be controlled 
through the optical phase of coherent light sources that drive the interaction, in such a way 
that the transition rates to final states and the dynamics at various stages of the process 
can be modified \cite{shapiro}. 

In a recent paper \cite{pra1}, a theoretical investigation on the quantum coherent control 
of the optical transient four-wave mixing of two intense phase-locked femtosecond laser 
pulses of central angular frequencies $\omega$ and $3\omega$ 
propagating in a two-level atom (TLA) was reported. It was shown how the nonlinear ($\chi^{(3)}$) 
coupling to the anti-Stokes Raman field at frequency $5\omega$ depends critically on the initial 
relative phase $\phi$ of the propagating pulses. In Ref. \cite{pra1}, the study was centered
to intense pulses in the visible and ultraviolet spectral regions,
with frequencies at resonance or lower than the atomic transition.
The phenomena observed in \cite{pra1}, however, can be scaled to various 
laser and material parameters.
In the infrared spectral range, for instance, experiments on ultrafast molecular 
dynamics are frequently performed by 
help of two-color pump probe nonlinear spectroscopy techniques. In this type of studies,
due to the high intensities inherent to ultrashort (subpicosecond) pulses, 
nonlinear effects such as stimulated Raman processes may become important 
and are often utilized as a complementary tool to gain information \cite{IRspectroscopy}.

In this Letter, we address the conditions for quantum 
coherent control of transient four-wave mixing interactions 
with two phase-locked $\omega$-$3\omega$ femtosecond 
laser pulses in the mid-IR spectral region. 
Our purpose is to reveal the basic physics for a strict two-level medium
and to this end the Maxwell-Bloch TLA will be considered.
We will study pulses with a duration of 300 fs 
(with a spectral width of $\approx 35$ cm$^{-1}$) and peak intensities 
as $\sim$10$^8$ W/cm$^2$, which are typically used in infrared nonlinear 
spectroscopy experiments \cite{IRspectroscopy}. We will examine the influence of frequency 
detuning by considering the field at $3\omega$ above resonance with respect to 
the atomic transition, something that was not considered in \cite{pra1}. 
Under these conditions, we will show that the quantum interferences leading to
a dominating efficiency of the nonlinear coupling to the anti-Stokes Raman component 
are governed by the ac Stark shift induced in the system.  
Furthermore, we will demonstrate that the relative spectral amplitude of the anti-Stokes 
fields produced by phase-locked pulses cancels
for frequency detunings that compensate the ac Stark effect.
This fundamental interference effect
 might be the basis for a novel ultrafast spectroscopy tool based on coherent control, 
which we name coherent ac Stark nonlinear spectroscopy (CSNS) for use later below. 

\begin{figure}[]
\begin{center}
\includegraphics[scale=0.40,clip=true,angle=270]{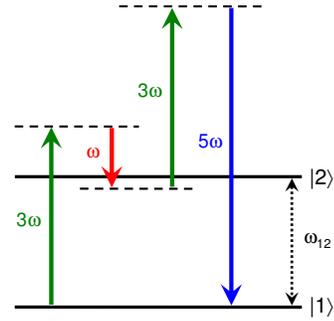}
\caption{\label{fig1} 
(Color online) Schematic energy level diagram. The resonance wavelength is 
considered in the mid-IR at $\lambda_{(\omega_{12})}=3000$ nm}
\end{center}
\end{figure}

The pulse propagation is modeled by means of the Maxwell-Bloch equations
beyond the rotating-wave approximation, 
allowing the resonant as well as the non-resonant regimes of the system 
to be described \cite{shapiro}. The equations are written as 
\begin{eqnarray}\label{MBS}
\frac{\partial{H}}{\partial{t}}&=&-\frac{1}{\mu_0}\frac{\partial{E}}{\partial{z}}, \nonumber
\end{eqnarray}
\begin{eqnarray}
\frac{\partial{E}}{\partial{t}}&=&-\frac{1}{\epsilon_0}\frac{\partial{H}}{\partial{z}}-
\frac{N_{at}\mu}{\epsilon_0 T_2}(\rho_1-T_2\omega_{12}\rho_2), \nonumber \\
\frac{\partial{\rho_1}}{\partial{t}}&=&-\frac{1}{T_2}\rho_1 +\omega_{12}\rho_2, \\
\frac{\partial{\rho_2}}{\partial{t}}&=&-\frac{1}{T_2}\rho_2 +\frac{2\mu}{\hbar}E\rho_3-\omega_{12}\rho_1, \nonumber \\
\frac{\partial{\rho_3}}{\partial{t}}&=&-\frac{1}{T_1}(\rho_3-\rho_{30})-\frac{2\mu}{\hbar}E\rho_2, \nonumber 
\end{eqnarray}
where $H(z,t)$ and $E(z,t)$ represent the magnetic and electric fields
propagating along the $z$ direction, respectively,
$\mu_0$ and $\epsilon_0$ are the magnetic permeability and 
electric permittivity of free space,  respectively,
$N_{at}=2\times10^{24}$ m$^{-3}$ is the density of polarizable atoms,
$\mu=4.2\times 10^{-29}$ Cm is the effective dipole coupling coefficient,
$T_1=T_2=1$ ps are the excited-state lifetime
and dephasing time, respectively, $\rho_1$ and $\rho_2$ are the real 
and imaginary components of the polarization,
and $\omega_{12}$ is the transition resonance angular frequency of 
the two level medium, considered in the present simulations in the mid-IR region
at 3000 nm (see Fig. \ref{fig1}).
The population difference is $\rho_3$, and $\rho_{30}$ represents its initial value.
An hyperbolic secant two-color pulse that can be expressed as 
\begin{eqnarray}
&E(t)=E_\omega(t)+E_{3\omega}(t)= E_0sech((t-t_0)/t_p)\times &  \nonumber \\
&\left[cos(\omega (t-t_0))+cos(3\omega (t-t_0)+\phi)\right]& \label{EINJ}
\end{eqnarray}
is externally injected to the system.
The peak input electric field amplitude $E_0$
is chosen the same for both pulses and results in an intensity 
of $4.0\times 10^8$ W/cm$^2$.
The duration of the pulses is given by $t_p=\tau_p /1.763$, 
with $\tau_p=300$ fs being the full width at half maximum (FWHM) of the
pulse intensity envelope. $t_0$ gives the offset position of the
pulse center at $t=0$, and it is the reference value for the 
phase of the pulses. The central angular frequencies of the 
pulses are $\omega$ and $3\omega$, and $\phi$ is the relative phase. 
The propagating system has been resolved numerically
by means of a standard finite difference time domain method described 
elsewhere \cite{pra1}.

\begin{figure}[]
\begin{center}
\includegraphics[scale=0.65,clip=true,angle=270]{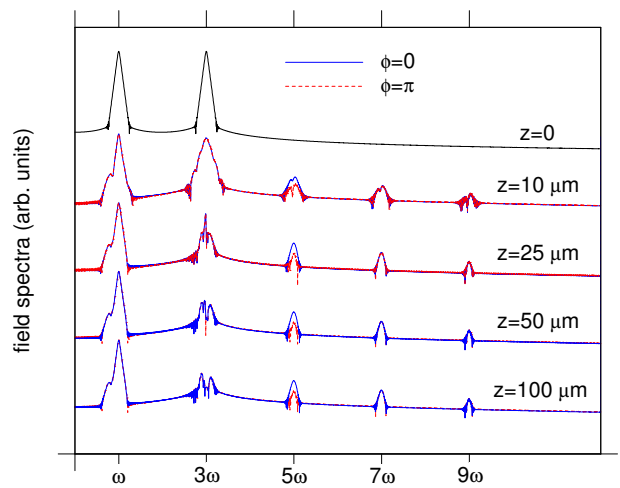}
\caption{\label{fig2} 
(Color online) Spectra of the total field at different propagation lengths as indicated.
In the case shown, the central pulse frequency $3\omega$ is 
in resonance with the atomic transition. Each spectrum is plotted in logarithmic scale.
The relative spectral amplitude of the anti-Stokes $5\omega$ components is 
shown quantitatively for $z=25$ $\mu$m in Fig. \ref{fig3} (case with $\eta=9.0$).}
\end{center}
\end{figure}
\begin{figure*}[]
\begin{center}
\includegraphics[scale=0.65,clip=true,angle=270]{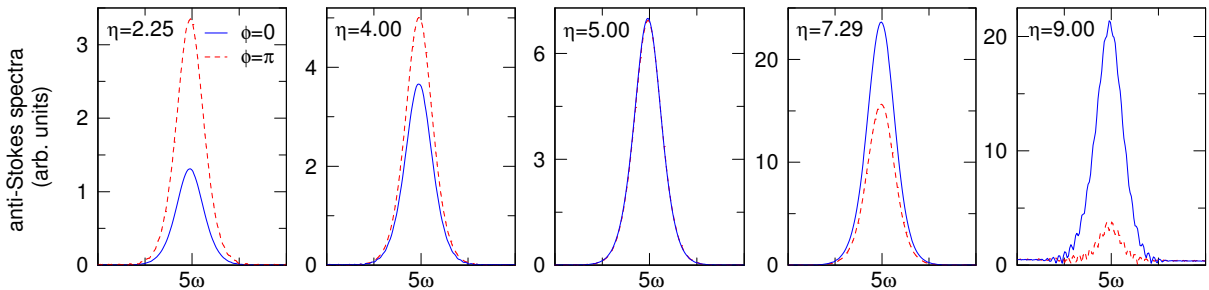}
\caption{\label{fig3} 
(Color online) Spectra for differents values of $\eta$ showing the anti-Stokes $5\omega$ frequency component of the
field at the propagation distance $z=25$ $\mu$m, for $\phi=0$ (solid lines) 
and $\phi=\pi$ (dashed lines). At $\eta=5$ (center plot) the dependence on the relative phase disappears.}
\end{center}
\end{figure*}

Figure 2 shows the field spectra at different propagation lengths
in the case that the central angular frequency of the field $E_{3\omega}$ is 
at resonance with the atomic transition ($3\omega=\omega_{12}$). 
The spectrum on the top is for the initial pulses.
The succeeding plots show the evolution of the spectrum as the pulses propagate
through the medium. In this case, we can observe the effect of the absorption of the pulse 
at $3\omega$, together with the appearance of other 
spectral contributions at $5\omega$, $7\omega$, 
and $9\omega$, which are produced as a result of the coupling of the fields through 
the third order nonlinearity ($\chi^{(3)}$) of the medium. 
It is clear that the conversion to the anti-Stokes Raman component ($5\omega$) 
depends on the relative phase between the pulses. For the parameter values
of the results shown in Fig. \ref{fig2}, 
the coupling to the fifth harmonic ($5\omega$) is more 
efficient for $\phi=0$ than for $\phi=\pi$,
an scenario that was already reported in Ref. \cite{pra1} for a transition 
in the visible region.
Hence here we confirm that observed in Ref. \cite{pra1} for a different system,
particularly, considering a mid-IR atomic transition. It is important to note  
that this phase dependence effect involves the anti-Stokes Raman component ($5\omega$) only, 
not the $7\omega$ nor the $9\omega$ spectral components, 
which remain insensitive to the initial relative phase of the pulses.  

We now turn to the study of the frequency detuning of the fields with respect to the
atomic transition. We will show that there is a central pulse frequency $\omega$ at which 
the relative phase dependence of the coupling to the anti-Stokes Raman component,
which has been previously discussed in Fig. \ref{fig2}, disappears.
We will observe that this effect can occur because the
ac Stark shifts produced in the medium by the fields $E_{\omega}$ and $E_{3\omega}$ can be 
compensated when $\omega<\omega_{12}<3\omega$.  
We will then conclude that the ac Stark shift in the medium governs 
the relative phase dependence of the coherent four-wave coupling. 

Indeed, the ac Stark frequency shift $\Delta \omega$ produced by a field of frequency $\omega$
which is not near resonance in a transition of frequency $\omega_{12}$  
can be expressed as \cite{acStark}:
\begin{eqnarray}\label{eq33}
\Delta\omega&=&-\frac{\omega_{12}-\omega}{2} \\ \nonumber
&&\pm\frac{1}{2}\left[(\omega_{12}-\omega)^2+4\beta^2(1+\frac{\omega_{12}-\omega}{\omega_{12}+\omega}) \right]^{1/2}, 
\end{eqnarray}
where $+$ is for $\omega<\omega_{12}$ and $-$ is for $\omega\gtrsim\omega_{12}$,
$\beta=\mu|E|/(2\hbar)$, with $|E|$ being the amplitude of the field, $\omega_{12}$ the
frequency of the atomic transition, and $\omega$ the field angular frequency. The
conditions $\beta/\omega <<1$ and $\beta/\omega_{12}<<1$ must be met for 
Eq. \ref{eq33} to be valid \cite{acStark}.
Far from resonance, where
\begin{eqnarray}\label{condeq36}
\frac{(\omega_{12}-\omega)^2}{\beta^2}&>>&\frac{8\omega_{12}}{\omega_{12}+\omega},
\end{eqnarray}
Eq. (\ref{eq33}) becomes
\begin{eqnarray}\label{acStarkshift}
\Delta\omega\approx \beta^2\left[ \frac{1}{\omega_{12}-\omega}+\frac{1}{\omega_{12}+\omega}\right].
\end{eqnarray}
Note that the last term inside the brackets in Eq. (\ref{acStarkshift})
is important far from resonance, where the rotating wave approximation does not apply. 
Clearly from Eq. (\ref{acStarkshift}), when $\omega>\omega_{12}$ we have $\Delta\omega<0$,  
and the separation in frequency of the states ($\omega_{12}+\Delta\omega$) 
appears to be less than in the absence of the field.
Contrarily, for $\omega<\omega_{12}$, the ac Stark frequency shift is positive. 

In the case of the present investigation, 
the main contributions to the ac Stark effect come from the two components
of the propagating pulses $E_\omega$ and $E_{3\omega}$, which have central angular frequencies $\omega$ and $3\omega$, respectively,
and the conditions for Eq. (\ref{acStarkshift}) are clearly met for the parameter values
of our study. Therefore, requiring that the combined ac Stark 
effect is null
\begin{eqnarray}\label{condition}
0=\frac{1}{\omega_{12}-\omega}+\frac{1}{\omega_{12}+\omega}+\frac{1}{\omega_{12}-3\omega}+\frac{1}{\omega_{12}+3\omega},&&
\end{eqnarray}
we obtain $\omega=\omega_{12}/\sqrt{5}$. 

We will next show that the ac Stark cancellation frequency $\omega=\omega_{12}/\sqrt{5}$
is indeed observed with accuracy from numerical simulations.
In Figure \ref{fig3}, the spectral amplitude of the anti-Stokes $5\omega$ component is shown for different
values of the detuning between the fields and the atomic transition at a propagation length such as $z=25 \mu$m. 
It is useful to define the parameter $\eta=(\omega_{12}/\omega)^2$, which sets the value of 
the detuning of the $E_{\omega}$ and $E_{3\omega}$ fields. The spectra corresponding to $\phi=0$ is shown by
solid lines in Fig. \ref{fig3}, while the spectra for $\phi=\pi$ is represented by the dashed lines. Clearly,
there is a switch in the tendency to dominate the conversion to the $5\omega$ anti-Stokes component.
For $\eta < 5$, the ac Stark shift due to $E_\omega$ dominates over the shift induced
by $E_{3\omega}$, and therefore the resulting separation of the states due to the combined Stark effect 
appears to be larger than in the absence of fields. 
In this situation, the coupling to the $5\omega$ anti-Stokes
field is enhanced for $\phi=\pi$,
as it can be observed in Fig. \ref{fig3}, left plots. Contrarily,
for $\eta>5$, when the ac Stark shift due to $E_{3\omega}$ dominates over the shift induced
by $E_{\omega}$, the coupling to the $5\omega$ anti-Stokes Raman
component is enhanced for $\phi=0$ (right plots in Fig. \ref{fig3}). In this last case 
the resulting separation in frequency of the states due to the total Stark effect 
is less than in the absence of fields. Furthermore, as shown in Fig. \ref{fig3}, 
the switching occurs at $\eta=5$, where the total ac Stark shift is cancelled,
 as expected from Eq. \ref{condition}. Note that for 
the numerical simulations to agree with the prediction of the analytical theory, the expression for the Stark effect beyond
the rotating wave approximation needs to be considered (see Eq. (33) in Ref. \cite{acStark}).  

\begin{figure}[]
\begin{center}
\includegraphics[scale=0.65,clip=true,angle=0]{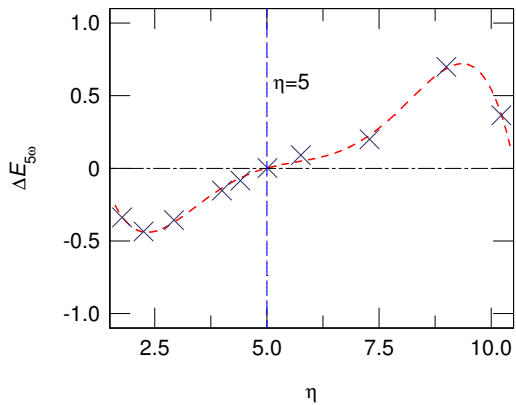}
\caption{\label{fig4} 
(Color online) Relative spectral amplitude of the anti-stokes Raman fields [as defined in Eq. (\ref{difampli})]
as a function of the parameter $\eta$ for $z=25 \mu$m. 
The vertical dashed line is $\eta=5$ as indicated. The dashed curve is a guide to the eye.}
\end{center}
\end{figure}

We now look at the relative spectral amplitude of the anti-stokes Raman fields, which we define as 
\begin{eqnarray}\label{difampli}
\Delta E_{5\omega}(z)&=&\frac{E^0_{5\omega}(z)-E^\pi_{5\omega}(z)}{E^0_{5\omega}(z)+E^\pi_{5\omega}(z)},
\end{eqnarray}
with $E^0_{5\omega}(z)$ being the spectral amplitude at the generated anti-Stokes ($5\omega$) frequency 
in the case that the initial relative phase is $\phi=0$ (see Fig. \ref{fig3}), and $E^\pi_{5\omega}(z)$ being
the spectral amplitude at $5\omega$ in the case that the initial relative phase between the pulses is $\phi=\pi$. Figure \ref{fig4}
shows the results obtained from the numerical simulations for $z=25 \mu$m. Relative differences in amplitudes 
as $|\Delta E_{5\omega}| \gtrsim 0.5$ can readily be produced in the cases considered in our simulations. 
Moreover, we have checked that the relative phase dependence  
at $\eta=5.0$ remains null for propagation distances as long as $z=100 \mu$m.     

We have therefore demonstrated that considering the coherent propagation of two-color phase-locked femtosecond pulses in 
a two-level medium, with central angular frequencies $\omega$ and $3\omega$, one can find an angular frequency
$\omega$ at which the ac Stark effect produced by the propagating pulses is cancelled. At this frequency value,
which requires the pulse with central frequency $3\omega$ to be above resonance with 
respect to the atomic transition, the phase dependence of the transient four-wave coupling
through the $\chi^{(3)}$ nonlinear susceptibility of the medium disappears. This is a fundamental 
quantum interference effect that to the best of our knowledge has not been 
reported before. 

In the present Letter, we have considered atomic frequencies which lie in 
the mid-IR region, which are of interest for applications e.g. in infrared nonlinear 
spectroscopic techniques. It has to be stressed however 
that the phenomena that we report can be 
scaled to several material and pulse parameters.
We can hence imagine straightforward applications for nonlinear ultrafast 
spectroscopy techniques based on the coherent control of subpicosecond 
two-color $\omega$-$3\omega$ propagating laser pulses. 
Indeed, the production of phase related $\omega$-$3\omega$ pulses is frequently 
accomplished by some frequency trippling mechanism with the subsequent 
variation of the phase of one of the pulses in order to obtain experimental 
control over the relative phase. Although our analysis has obviously been simplified 
by considering two well isolated levels as a first approach, the switching effect discussed here
is of a fundamental level, and in that sense it should be observed experimentally in particular media
where the two-level approximation is met. For instance,
some gaseous atoms have well isolated resonances  (as e.g. rubidium in the visible region). 
Also, the two-level approximation can be used for studying coherent 
effects in materials with a broad distribution of transitions, such as inhomogeneously broadened 
resonance lines in gases and in condensed matter, and even for inhomogeneous 
quasicontinuous energy bands as in semiconductors \cite{twolevelatoms,twolevelexperiments}.
CSNS can hence provide information on the transitions 
being probed by measuring the relative spectral amplitude of the 
anti-Stokes Raman fields with a simple scan of the laser frequency.  
The ac Stark mediated coherent control scenario reported in this Letter has therefore a broad interest and
may be the basis for future studies on more complex systems.     

Support from the {\it Programa Ram\'on y Cajal} of the Spanish Ministry of Science and Technology and from
project FIS2004-02587 is acknowledged.

{}
\end{document}